\documentclass[12pt,showpacs,preprintnumbers,amsmath,amssymb]{revtex4}
\usepackage{graphicx}
\begin{document}

\newcommand{\pderiv}[2]{\frac{\partial #1}{\partial #2}}
\newcommand{\deriv}[2]{\frac{d #1}{d #2}}
\newcommand{\eq}[1]{Eq.~(\ref{#1})}  

\title{Spin-Glass Attractor on Tridimensional Hierarchical Lattices in 
the Presence of an External Magnetic Field}

\vskip \baselineskip

\author{Octavio R. Salmon}
\thanks{E-mail address: octavior@cbpf.br}

\author{Fernando D. Nobre}
\thanks{Corresponding author: E-mail address: fdnobre@cbpf.br}

\address{
Centro Brasileiro de Pesquisas F\'{\i}sicas \\
Rua Xavier Sigaud 150 \\
22290-180 \hspace{5mm} Rio de Janeiro - RJ \hspace{5mm} Brazil}

\date{\today}


\begin{abstract}
\noindent
A nearest-neighbor-interaction Ising spin glass, in the presence of an
external magnetic field, is studied on different hierarchical lattices 
that approach the cubic lattice. The magnetic field is considered as
uniform, or random (following either a bimodal or a Gaussian probability
distribution). In all cases, a spin-glass attractor is found, in the plane
magnetic field versus temperature, associated with a low-temperature phase.
The physical consequences of this attractor are discussed, in view of the
present scenario of the spin-glass problem.  

\vskip \baselineskip


\end{abstract}
\maketitle

\newpage

\section{Introduction}

Short-range-interaction spin-glass (SG) models 
\cite{dotsenkobook,nishimoribook,youngbook,fischerhertz} have raised a lot
of controversies in the last decades. From the theoretical point of view, 
the case of Ising spins represents the most convenient to be investigated,
in such a way that a large effort has been dedicated to the understanding
of the Ising SG model. The majority of works concentrated
on three-dimensional Ising SG models, for which, besides its physical
realizations, it is generally accepted
nowadays that a SG phase occurs at finite temperatures
\cite{southernyoung,bhattyoung85,ogielski85,braymoore87,bhattyoung88,%
parisireview,young04,young06,jorg}. 

Hierarchical lattices have been a very useful tool for the study of
SG models
\cite{southernyoung,braymoore87,banavar87,banavar88,niflehilhorst92,%
nobre98,drossel98,nogueira99,curado99,%
marinaricomment,drosselreply,nobre00,drossel00,nobre01,nobre03},
essentially due to the possibility of carrying out exact calculations, or
performing relatively low-time-consuming numerical computations.
Apart from a few exceptions \cite{nobre98,nobre00,nobre01,nobre03}, most of
the hierarchical lattices considered so far belong to the 
Migdal-Kadanoff (MK) family. Taking into account the significant reduction
of efforts in the investigation of these models, some of the results 
obtained are quite impressive:
(i) The lower critical dimension, below which
the SG phase transition occurs at zero temperature, which was the
object of a lot of controversy about three decades ago, 
was correctly estimated
within a MK renormalization-group (RG) approach \cite{southernyoung}, 
almost a decade before the consensus that this quantity should be greater
than 2, but smaller than 3, through different numerical approaches, like 
numerical simulations \cite{bhattyoung85,ogielski85,bhattyoung88} and 
zero-temperature domain-wall arguments \cite{braymoore87}; 
(ii) The critical-temperature estimates of Ref.~\cite{southernyoung}
for an Ising SG on a MK hierarchical lattice of fractal dimension $D=3$, 
with symmetric distributions, 
are $(k_{B}T_{c} /J) = 1.05 \pm 0.02$ \ ($\pm J$ distribution) and 
$(k_{B}T_{c} /J) = 0.88 \pm 0.02$ \ (Gaussian distribution of width $J$). 
The most recent Monte Carlo simulations on a cubic 
lattice \cite{young06} yield 
$(k_{B}T_{c} /J) = 1.120 \pm 0.004$ in the first case, whereas for the later, 
$(k_{B}T_{c} /J) = 0.951 \pm 0.009$, leading to relative discrepancies of 
about $4 \% $ when compared with the results of 
Ref.~\cite{southernyoung}, taking into account the error bars; 
(iii) A zero-temperature analysis of a Gaussian Ising SG, on a 
special hierarchical lattice with fractal dimension
$D=2$ \cite{nobre98}, yielded an estimate for the 
stiffness exponent $y$ ($y=-1/\nu$, where
$\nu$ is the exponent associated with the divergence of the
correlation length at zero temperature) in agreement with those
obtained from other, more time-consuming, numerical approaches; 
(iv) The same hierachical lattice of Ref.~\cite{nobre98}, mentioned above,
produced a precise ferromagnetic-paramagnetic 
critical frontier for the $\pm J$ Ising SG model \cite{nobre01}. 

A major question in the SG problem nowadays concerns the applicability of
some  results from the mean-field solution for short-range-interaction
systems. In particular, whether the SG phase is properly described by an
infinite number of order parameters (i.e., an order-parameter function
\cite{parisirsb}), manifesting the property of replica-symmetry  
breaking (RSB) \cite{parisireview}; moreover, if an Ising SG, in the
presence of an external magnetic field, exhibits the Almeida-Thouless (AT)
line \cite{at}, which separates a low-temperature region characterized by
RSB, from a high-temperature one, described in terms of a single order
parameter, along which the replica-symmetric solution holds. 
Numerical simulations for nearest-neighbor-interaction Ising SGs are 
always hard to
be performed, since equilibration becomes difficult for large system sizes
and low temperatures. However, in spite of the small lattice sizes
considered, there are evidences from Monte Carlo simulations that a
critical line in the presence of a field exists in $d=4$ \cite{parisi4d}, 
but not in $d=3$ \cite{young04,jorg}. Furthermore, a zero-temperature 
analysis of the energy landscape in the case $d=3$ is compatible with a 
transition from the SG to the paramagnetic phase for a finite critical 
field, although the possibility of a critical field equal to zero was not 
excluded \cite{krzakala01}. 
However, it is possible to have a critical frontier separating the SG and 
paramagnetic phases, for low-dimensional
short-range-interaction Ising SGs in the 
presence of a magnetic field, which is not an AT-like line. In order to
ensure that this critical line represents a true AT line, one should also 
verify evidences of RSB below such a frontier. This possibility 
would correspond to an
``intermediate'' scenario \cite{krzakala01}, between the mean-field 
RSB solution and the much simpler droplet picture 
\cite{dotsenkobook,nishimoribook,youngbook}. In this case, one would
expect that RSB effects should appear below this line, at some finite
dimension $d$, contrary to claims of the droplet model, according to
which the AT line should occur only in the limit of infinite dimensions.

One of the advantages for the study of SGs on hierarchical lattices 
is that one does not go through equilibration difficulties. In view of
this, the $D=3$ MK hierarchical lattice has been used also for an 
investigation of RSB in the low-temperature phase on an Ising SG without
a magnetic field \cite{drossel98,marinaricomment,drosselreply,drossel00}.
For temperatures in the range $0.7 \ T_{c}<T<T_{c}$, a picture showing
characteristics of RSB was observed, although for lower temperatures the
results agree with the simpler, replica-symmetric scenario. 

To our knowledge, short-range-interaction SGs in the presence of an
external magnetic field have never been investigated on hierarchical
lattices. In the present work, we study an Ising SG model, in the presence
of different types of external magnetic fields, on three
hierarchical lattices (two of them with a fractal dimension $D=3$ and
another one characterized by $D \approx 3.58$). In the next section we
define the model, the hierarchical lattices, and the numerical procedure.
In section III we present and discuss our results. 
   
\section{The Model and Numerical Procedure}   
   
\begin{figure}[t]
\begin{center}
\includegraphics[width=0.30\textwidth,angle=0]{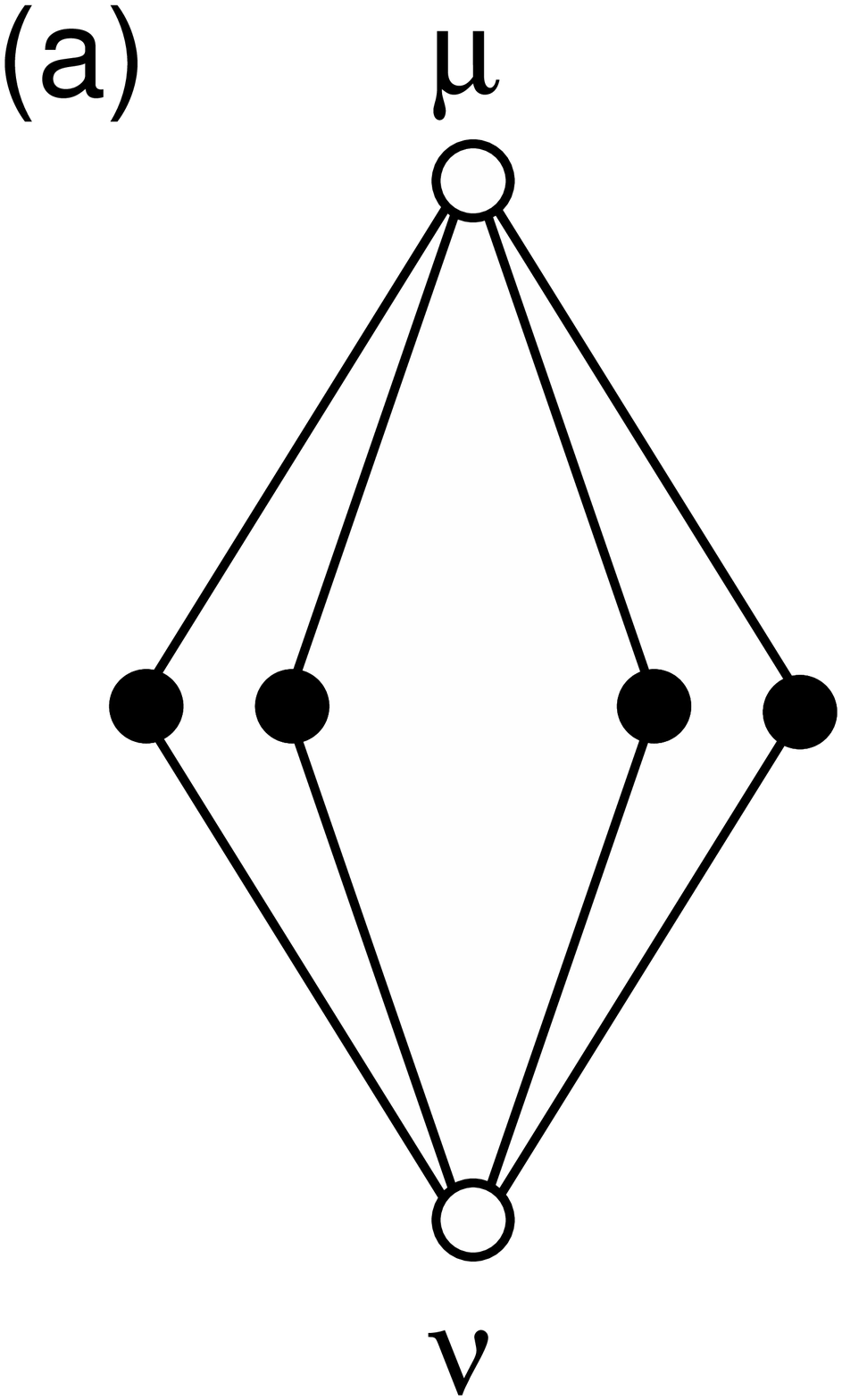}
\hspace{0.5cm}
\includegraphics[width=0.30\textwidth,angle=0]{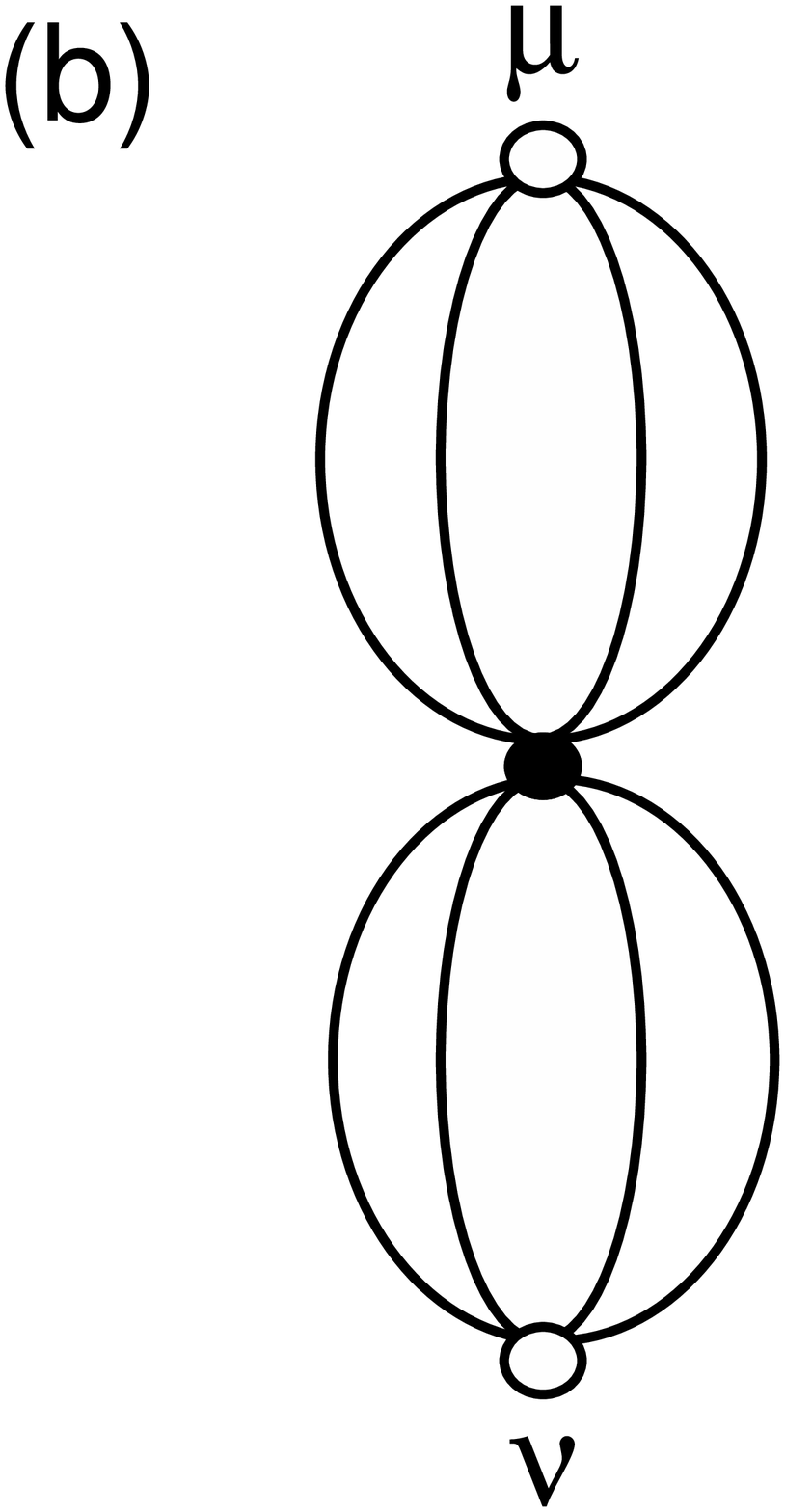}
\hspace{0.5cm}
\includegraphics[width=0.30\textwidth,angle=0]{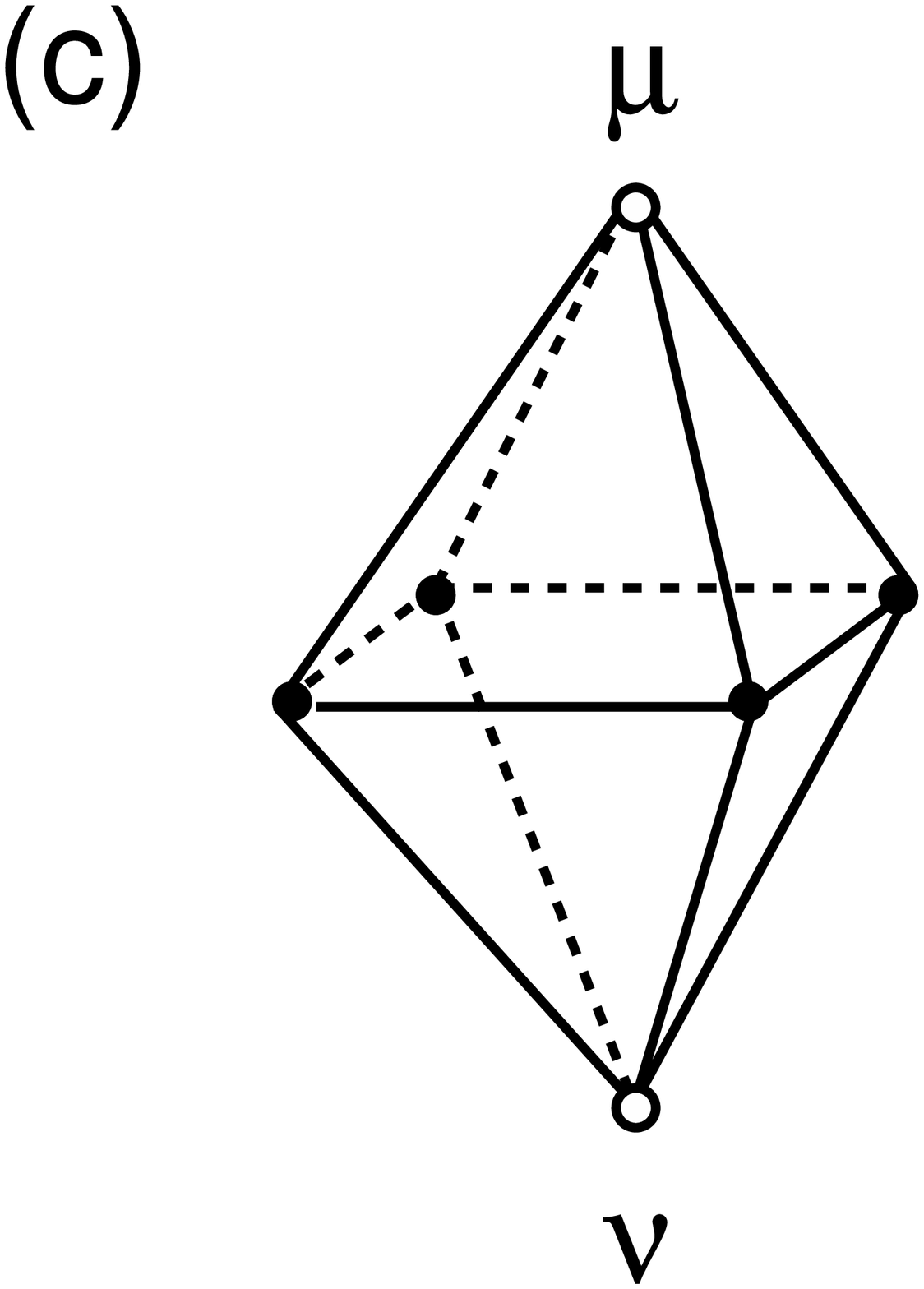}
\end{center}
\protect\caption{Basic cells of three hierarchical lattices that approach
the cubic lattice. 
(a) The MK cell of fractal dimension $D=3$ (usually called of diamond cell); 
(b) The dual of the diamond cell (fractal dimension $D=3$);
(c) The tridimensional Wheatstone-Bridge cell (fractal dimension
$D \approx 3.58$). The empty circles represent the external sites, whereas
the black circles are sites to be decimated in the renormalization process.}
\label{fig:hlattices}
\end{figure}

Herein, we study an Ising SG, 
in the presence of an external magnetic field, 

\begin{equation}
\label{eq:hamiltonian}
{\cal H} = - \displaystyle\sum_{<ij>}J_{ij}S_{i}S_{j}-\sum_{i}h_{i}S_{i}
 \quad (S_{i}= \pm 1), 
\end{equation}

\vskip \baselineskip
\noindent
where $\{ J_{ij} \} $ denote random couplings between two spins located at
nearest-neighboring sites $i$ and $j$ of a given hierarchical lattice, 
following a symmetric Gaussian probability distribution,

\begin{equation} 
\label{eq:pjij}
P(J_{ij})=\frac{1}{\sqrt{2\pi J^{2}}}
\ \exp\left(-\frac{J_{ij}^{2}}{2J^{2}} \right)~.   
\end{equation}

\vskip \baselineskip
\noindent
For the magnetic fields we consider three different cases, namely, 

\begin{equation}
\label{eq:unifield}
P(h_{i})=\delta(h_{i}-H_{0}) \quad ({\rm uniform \ field}),
\end{equation}

\begin{equation}
\label{eq:bimfield}
P(h_{i})=\frac{1}{2} \ \delta(h_{i}-H_{0})
+\frac{1}{2} \ \delta(h_{i}+H_{0})
\quad ({\rm symmetric \ bimodal \ distribution}),
\end{equation}

\begin{equation} 
\label{eq:gaussfield}
P(h_{i})= \frac{1}{\sqrt{2\pi\sigma^{2}}}
\ \exp\left[-\frac{(h_{i}-H_{0})^{2}}{2\sigma^{2}} \right]
\quad ({\rm Gaussian \ distribution}). 
\end{equation}

\vskip \baselineskip
\noindent
The Hamiltonian of \eq{eq:hamiltonian} will be investigated on three different 
hierarchical lattices that approach the cubic lattice. These lattices are
generated by starting the process from the 0th level of the lattice-generation 
hierarchy, 
with a single bond joining the external sites (denoted by $\mu$ and $\nu$).
Then, in each iteration step one replaces a single bond by a unit cell,
like the ones 
shown in Fig.~\ref{fig:hlattices}, in such a way that in its first 
hierarchy, each lattice is
represented by a unit cell; the hierarchical lattice is constructed
up to a given ${\cal N}$-th hierarchy (${\cal N} \gg 1$).  The cells in 
Figs.~\ref{fig:hlattices}(a) and (b) are considered as dual to one
another, and their results for critical temperatures 
usually represent lower
and upper limits, respectively, with respect to the correct value
of the cubic lattice \cite{melrose83}. The cell 
of Fig.~\ref{fig:hlattices}(c) is a three-dimensional 
Wheatstone Bridge cell, and to our knowledge, it 
has never been used in the study or random magnetic models.

The RG procedure works in the inverse way of the lattice
generation, i.e., through a decimation of the internal sites of 
a given cell, leading to renormalized quantities associated with the
external sites. Defining the dimensionless couplings and fields,
$K_{ij}=\beta J_{ij}$ , $H_{i}=\beta h_{i}$ \ $[\beta=1/(k_{B}T)]$, 
the corresponding RG equations may be written in the
general form (see Ref.~\cite{rosascoutinho04} for  
the explicit form of these equations for the MK
cell of Fig.~\ref{fig:hlattices}(a)),

\begin{equation}
\label{eq:kp} 
K^{'}_{\mu \nu}=\frac{1}{4}
\log\left(\frac{Z_{--}Z_{++}}{Z_{-+}Z_{+-}}\right)~,   
\end{equation}

\begin{equation}
\label{eq:hmup} 
H^{'}_{\mu}=\frac{1}{4} 
\log\left(\frac{Z_{++}Z_{+-}}{Z_{--}Z_{-+}}\right)~,   
\end{equation}

\begin{equation}
\label{eq:hnup} 
H^{'}_{\nu}=\frac{1}{4} 
\log\left(\frac{Z_{++}Z_{-+}}{Z_{--}Z_{+-}}\right)~,   
\end{equation}

\vskip \baselineskip
\noindent
where $Z_{S_{\mu},S_{\nu}}$ represent partition functions of a given cell
with the external spins kept fixed $(S_{\mu},S_{\nu}=\pm 1)$,

\begin{equation} 
Z_{S_{\mu},S_{\nu}} = {\rm Tr}_{\{S_{i} \ (i\neq \mu,\nu)\}}
\ [\exp(-\beta {\cal H})]~.   
\end{equation}

\vskip \baselineskip
\noindent
It should be noticed that the Hamiltonian associated with any of the cells
of Fig.~\ref{fig:hlattices} may be split into  
${\cal H} = {\cal H}^{'} + H_{\mu}S_{\mu} + H_{\nu}S_{\nu}$, where 
${\cal H}^{'}$ represents the Hamiltonian of the cell with
$H_{\mu}=H_{\nu}=0$. Therefore, the partition function 
$Z_{S_{\mu},S_{\nu}}$ may be rewritten as,

\begin{equation} 
Z_{S_{\mu},S_{\nu}} = \exp(H_{\mu}S_{\mu}+H_{\nu}S_{\nu}) 
{\cal{Z} }_{S_{\mu},S_{\nu}}~, 
\end{equation} 

\vskip \baselineskip
\noindent
where 

\begin{equation} 
{\cal Z}_{S_{\mu},S_{\nu}}  = {\rm Tr}_{\{S_{i} \ (i\neq \mu,\nu)\}}\ 
[\exp(-\beta {\cal H}^{'})]     
\end{equation}

\vskip \baselineskip
\noindent
represents the partition function of a given cell without considering
the contributions of the fields on its external sites. 
Accordingly, Eqs.~(\ref{eq:kp})--(\ref{eq:hnup}) take the following form, 

\begin{equation}
\label{eq:kp2} 
K^{'}_{\mu \nu}=\frac{1}{4}
\log\left(\frac{{\cal Z}_{--}{\cal Z}_{++}}
{{\cal Z}_{-+}{\cal Z}_{+-}}\right)~,   
\end{equation}

\begin{equation}
\label{eq:hmup2} 
H^{'}_{\mu}=H_{\mu}+\frac{1}{4} 
\log\left(\frac{{\cal Z}_{++}{\cal Z}_{+-}}
{{\cal Z}_{--}{\cal Z}_{-+}}\right)~,   
\end{equation}

\begin{equation}
\label{eq:hnup2} 
H^{'}_{\nu}=H_{\nu}+\frac{1}{4} 
\log\left(\frac{{\cal Z}_{++}{\cal Z}_{-+}}
{{\cal Z}_{--}{\cal Z}_{+-}}\right)~.   
\end{equation}

\vskip \baselineskip
\noindent
Thus, we clearly see through Eqs.~(\ref{eq:hmup2}) and (\ref{eq:hnup2})
that after the RG transformation, the  field on each remaining 
site depends on the previous field on this site plus a contribution 
due to the decimation of the inner spins in the cell. 

If at the beginning of the RG procedure
(${\cal N}$-th hierarchy), one uses the
probability distribution of \eq{eq:pjij} for the couplings, and one of the
probability distributions of Eqs.~(\ref{eq:unifield})--(\ref{eq:gaussfield}) 
for the fields, one has at this level, the average values,  
$<K_{ij}>=0$, $<H_{i}>=H_{0}$ (uniform field and Gaussian distribution),  
or $<H_{i}>=0$ (symmetric bimodal distribution). An important quantity to
be used herein is the ratio of associated widths,

\begin{equation} 
\label{eq:widthsratio}
r = \frac{\sigma_{K}}{\sigma_{H}}~; \quad 
\sigma_{K}=<(K_{ij}-<K_{ij}>)^{2}>^{1/2}~; \quad 
\sigma_{H}=<(H_{i}-<H_{j}>)^{2}>^{1/2}~, 
\end{equation}

\vskip \baselineskip
\noindent
which, at the ${\cal N}$-th level, 
is infinite for a uniform field (since $\sigma_{H}=0$ in this case), 
whereas 
$r=J/H_{0}$ (bimodal distribution) and $r=J/\sigma$ (Gaussian distribution
for the fields). However, as the renormalization procedure goes on (in
fact, right after the first RG transformation), 
the system of coupled equations that define the renormalization
[Eqs.~(\ref{eq:kp})--(\ref{eq:hnup})]
introduces correlations between the couplings and fields, and as 
a result of this, one has a joint probability distribution, 
$P(K_{ij},H_{i},H_{j})$, to be followed.

We have used the method proposed in Ref.~\cite{caomachta} to follow this
distribution numerically. This technique consists in generating a pool of 
$M$ triplets $\{K_{ij},H_{i},H_{j}\}$, initially chosen by generating random 
numbers according to the above distribution for the couplings, 
and one of the probability distributions for the fields.
An iteration consists in $M$
operations, where in each of them one picks randomly triplets from the
pool (each chosen triplet is assigned to a bond in one of the cells of 
Fig.~\ref{fig:hlattices}) in order to
generate the effective quantities of Eqs.~(\ref{eq:kp})--(\ref{eq:hnup})
that will define a triplet of the renormalized pool. 
At zero temperature
Eqs.~(\ref{eq:kp})--(\ref{eq:hnup}) become much simpler (see, e.g.,
Ref.~\cite{rosascoutinho04} for the explicit form of these equations 
for the MK cell of Fig.~\ref{fig:hlattices}(a)), 
and we have used these simplified forms 
to determine the zero-temperature critical points.
It is important to notice that in this procedure there may occur 
superpositions of random fields in given sites of the unit cells; whenever
this happens, we consider an arithmetic average of the superposed 
random fields. After each iteration, one calculates the lowest moments
associated with the couplings and fields, and in particular, the ratio
$r$ defined in \eq{eq:widthsratio}. 

Below, we present and discuss the results obtained from this formalism. 

\section{Results and Discussion}

For the results that follow, we have used $M=160000$, although we
checked that our results did not change (within the error bars) 
for larger pools of triplets. 
For all three hierarchical lattices considered, the first moments 
presented small fluctuations around their initial values 
[i.e., $<K_{ij}>=0$ and $<H_{i}>$ (either zero or not)] under successive 
renormalizations. Therefore, our critieria for the identification of the
attractors, and their associated phases, was based on the widths 
$\sigma_{K}$, $\sigma_{H}$, and their ratio $r$, which showed that the
behavior of $\sigma_{K}$ always prevailed over the one of $\sigma_{H}$. 

\begin{figure}[t]
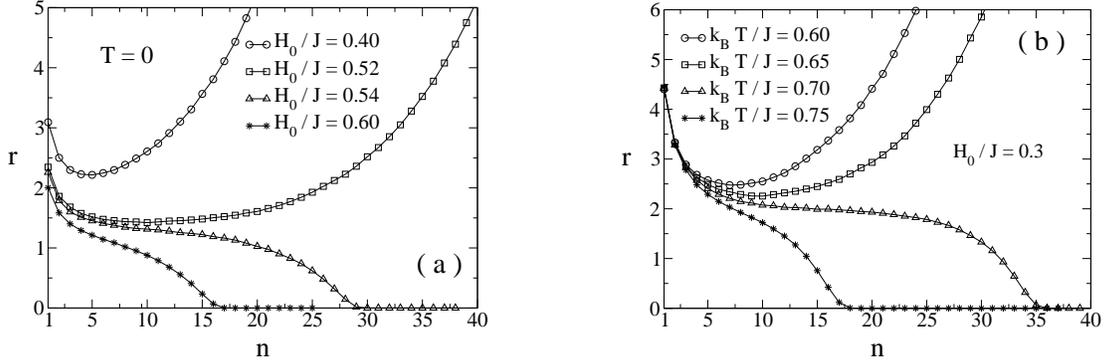

\begin{center}
\includegraphics[width=0.40\textwidth,angle=0]{fig2a.eps}
\hspace{1.5cm}
\includegraphics[width=0.40\textwidth,angle=0]{fig2b.eps}
\end{center}
\protect\caption{Evolution of the ratio $r$ with the RG
steps $n$, for an Ising SG on the hierarchical lattice defined by the unit
cell of Fig.~\ref{fig:hlattices}(a), in the presence of an initial uniform 
field $H_{0}$.
(a) Typical values of $H_{0}/J$ at zero temperature; the critical field is in
the range $0.52<(H_{0c}/J)<0.54$. (b) Typical temperatures for
$(H_{0}/J)=0.3$; the critical temperature is in the range
$0.65<(k_{B}T_{c}/J)<0.70$.}
\label{fig:ratior}
\end{figure}

We have found typically two distinct
behaviors for these quantities, under successive RG
transformations, which we associated with two
distinct phases, as described below.
(i) Paramagnetic ({\bf P}) phase: this occurs for sufficiently 
large temperatures and/or fields, where 
$\sigma_{K} \rightarrow 0$, $\sigma_{H} \rightarrow 0$, with
$r \rightarrow 0$. 
(ii) {\bf SG} phase: this phase appears for
low temperatures and fields, being characterized by 
$\sigma_{K} \rightarrow \infty$ and $r \rightarrow \infty$. In most of the
cases, we have found throughout this phase an increase on $\sigma_{H}$ as
well, but still keeping $r \rightarrow \infty$; however, in some situations
(essentially for symmetric field distributions), we have found that 
$\sigma_{K} \rightarrow \infty$ and $\sigma_{H} \rightarrow 0$.
The critical frontier separating these two phases
was considered as the one where 
the parameter $r$ changes very slowly. Typical behaviors of the ratio $r$,
under successive RG transformations, 
are illustrated in Fig.~\ref{fig:ratior} for an Ising SG on the 
hierarchical lattice defined by the unit
cell of Fig.~\ref{fig:hlattices}(a), in the presence of an initial uniform 
field $H_{0}$. In Fig.~\ref{fig:ratior}(a) we present this quantity for
different values of $H_{0}/J$ at zero temperature, whereas in 
Fig.~\ref{fig:ratior}(b), we have fixed the initial field [$(H_{0}/J)=0.3$]
and have varied the temperature; in both figures one sees that the ratio of
widths $r$ may present qualitatively distinct behaviors, 
depending on the initial parameters considered, signalling different 
attractors of the renormalization. 

\begin{figure}[t]
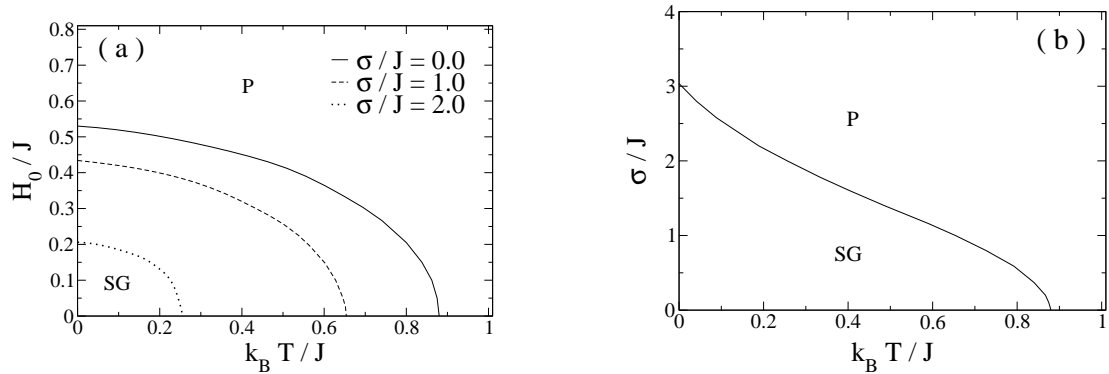

\begin{center}
\includegraphics[width=0.40\textwidth,angle=0]{fig3a.eps}
\hspace{1.5cm}
\includegraphics[width=0.40\textwidth,angle=0]{fig3b.eps}
\end{center}
\protect\caption{Phase diagrams for an Ising SG on the hierarchical 
lattice defined by the unit cell of Fig.~\ref{fig:hlattices}(a), 
in the presence of an initial random magnetic field following a Gaussian
probability distribution. (a) Plane $H_{0}$
versus temperature, for different values of $\sigma$. (b) Plane $\sigma$
versus temperature, for $H_{0}=0$. {\bf P} and {\bf SG} denote
the paramagnetic and spin-glass phases, respectively.} 
\label{fig:phasediaggauss}
\end{figure}

In Fig.~\ref{fig:phasediaggauss} we exhibit phase diagrams obtained from
the present RG approach for the Ising SG defined by 
Eqs.~(\ref{eq:hamiltonian}), (\ref{eq:pjij}), and (\ref{eq:gaussfield}),
on the diamond hierarchical lattice of Fig.~\ref{fig:hlattices}(a). 
In Fig.~\ref{fig:phasediaggauss}(a) we present phase diagrams in the
plane $H_{0}$ versus temperature for different choices of $\sigma$;
one notices that for increasing values of $\sigma$, 
the SG phase decreases; a similar effect has been observed in 
mean-field theory, where this phase is delimited by an AT line,
associated with 
RSB \cite{soares}. It is important to stress that the case 
$\sigma=0$ in Fig.~\ref{fig:phasediaggauss}(a) corresponds, within the
RG procedure, to an initial uniform field; therefore,
the uniform and Gaussian random fields produce a qualitatively similar 
critical frontier. 
In Fig.~\ref{fig:phasediaggauss}(b) we show the phase diagram in the 
plane $\sigma$ versus temperature, for the case of 
a symmetric Gaussian distribution for the fields ($H_{0}=0$), 
which displays also a SG phase for low temperatures. 

\begin{figure}[t]
\begin{center}
\includegraphics[width=0.60\textwidth,angle=0]{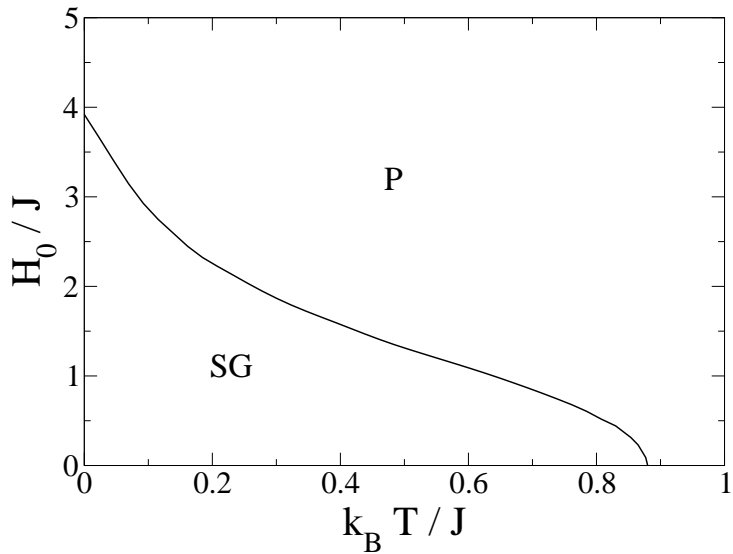}
\end{center}
\protect\caption{Phase diagram for an Ising SG on the hierarchical 
lattice defined by the unit cell of Fig.~\ref{fig:hlattices}(a), 
in the presence of an initial random magnetic field following a symmetric
bimodal probability distribution.}
\label{fig:phasediagbim}
\end{figure}

In Fig.~\ref{fig:phasediagbim} we exhibit the phase diagram 
for the present Ising SG model under the bimodal random 
field defined by Eq.~(\ref{eq:bimfield}). One notices that the critical
frontier of Fig.~\ref{fig:phasediagbim} presents a change of concavity for
low temperatures and so, it is qualitatively different from those of a 
uniform field [case $\sigma=0$ in Fig.~\ref{fig:phasediaggauss}(a)] 
and Gaussian random field [cases $\sigma>0$ 
in Fig.~\ref{fig:phasediaggauss}(a)]. 
Curiously, this critical frontier is very similar to the one of a
symmetric Gaussian distribution, shown in Fig.~\ref{fig:phasediaggauss}(b).
Analogous to the previous cases, shown in Fig.~\ref{fig:phasediaggauss}
(uniform and Gaussian random fields), the most important aspect,
i.e., the existence of the SG attractor, associated with a SG phase at
low temperatures, applies for the bimodal random field, as well. 

We have also investigated the present Ising SG model on the hierarchical
lattices of Figs.~\ref{fig:hlattices}(b) and \ref{fig:hlattices}(c); 
although the results are quantitatively different, 
the qualitative behaviors of the phase diagrams are the same as those 
shown above for the diamond hierarchical lattice. In Tables~I and II we
compare critical parameters, associated with the phase diagrams, for the
three hierarchical lattices investigated. In Table~I we present the 
values of the critical temperatures for $H_{0}=0$, in the case of the
uniform and bimodal distributions for the fields, whereas for the Gaussian
distribution, the results refer to the particular width $(\sigma/J)=1$. In
Table~II we present values of critical fields, at zero temperature; for a 
Gaussian distribution, our zero-temperature results correspond to the
cases, 
$(\sigma/J)=1$, or $H_{0}=0$ (cf., e.g., Fig.~\ref{fig:phasediaggauss}).
>From these tables one notices that the critical parameters for the
hierarchical lattices defined by the cells in 
Figs.~\ref{fig:hlattices}(a) and (b) are always below and above,
respectively, the ones of the hierarchical lattice of 
Fig.~\ref{fig:hlattices}(c). This confirms the current belief that
these lattices yield lower and upper limits for phase-diagram
critical parameters \cite{melrose83}; 
in addition to that, it suggests that the
Wheatstone-Bridge lattice of Fig.~\ref{fig:hlattices}(c) 
may lead to good approximations
in the study of three-dimensional SG systems. 
In fact, the corresponding critical-temperature estimate for $H_{0}=0$
(uniform and bimodal cases) shown in Table~I, $(k_{B}T_{c}/J)=0.980(1)$, is
in good agreement with the recent estimate from 
Monte Carlo simulations on a cubic lattice, 
$(k_{B}T_{c} /J) = 0.951(9)$ \cite{young06}, 
leading to a relative discrepancy of 
$2 \% $, taking into account the error bars. 

\begin{figure}[t]
\begin{center}
\includegraphics[width=0.60\textwidth,angle=0]{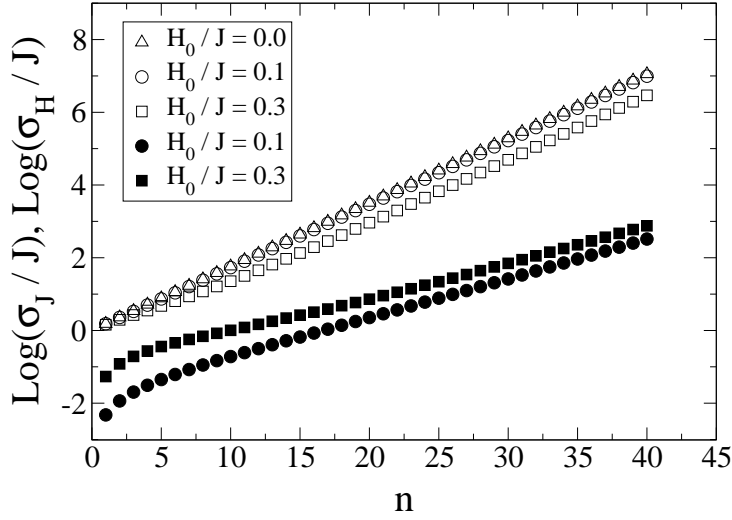}
\end{center}
\protect\caption{Evolution of the widths 
associated with the couplings $(\sigma_{J})$ (empty symbols) 
and fields $(\sigma_{H})$ (black symbols) with the RG steps $n$ for
typical values of an initial uniform field $H_{0}$. In the first case we
exhibit also the case $H_{0}=0$.} 
\label{fig:stiffness}
\end{figure}

The results above are reinforced by a zero-temperature analysis of the
evolution of the widths
associated with the couplings $(\sigma_{J})$ and fields $(\sigma_{H})$,
with respect to the RG steps $n$, 

\begin{equation} 
\label{eq:stiffness}
\sigma_{J} \sim b^{ny}~; \quad 
\sigma_{H} \sim b^{nu}~, 
\end{equation}

\vskip \baselineskip
\noindent
which are characterized by the exponents $y$ (usually known as the stiffness
exponent) and $u$; the scaling factor is $b=2$ for all three cells of 
Fig.~\ref{fig:hlattices}.
In Fig.~\ref{fig:stiffness} we illustrate the power-law behaviors 
of \eq{eq:stiffness}
for the case of an initial uniform 
field $H_{0}$ on a hierarchical lattice defined by the unit cell of 
Fig.~\ref{fig:hlattices}(a). The 
values of $H_{0}$ chosen in Fig.~\ref{fig:stiffness} correspond to
zero-temperature points in the spin-glass phase of the case $\sigma=0$ in  
Fig.~\ref{fig:phasediaggauss}(a). 
The scaling range (in the RG steps $n$) starts for small $n$ (typically
around $n=5$) for $0 \leq H_{0} \lesssim H_{0c}/2$, but 
for larger values of $H_{0}$
(essentially as one approaches the critical value $H_{0c}$) the scaling
forms of \eq{eq:stiffness} are only satisfied for higher ranges of the
iteration steps $n$. We have noticed that, in all cases, 
the exponents $y$ and $u$ are weakly dependent on
the initial value $H_{0}$ (as can be seen on the examples shown in 
Fig.~\ref{fig:stiffness}); as a consequence of this, we have assumed 
universality with
respect to $H_{0}$, for each of these exponents. 
Our estimates, for the case of an initial uniform
field $H_{0}$, are presented in
Table~III for each of the
hierarchical lattices of Figs.~\ref{fig:hlattices}; 
the error bars take into account slight variations in these
exponents throughout the interval $0 \leq H_{0} \leq H_{0c}$. The fact that
$y>u$ supports the existence of a low-temperature SG phase, based on the
criterion described above for the quantity $r$ of \eq{eq:widthsratio}. 

At this point, it is important to stress that for the families of
hierarchical lattices associated with the cells in 
Figs.~\ref{fig:hlattices}(a) and (b), for which the fractal dimension may
be changed easily by varying the number of parallel paths, we have also
investigated lattices with fractal dimensions $D<3$ (more specifically, those
characterized by three parallel paths, for which $D \approx 2.585$). In
these cases, we did not find any evidence of a SG attractor, suggesting
that for these lattices one has a lower critical dimension,
$2.585<D_{l}<3$. 
Taking into account the above-mentioned properties of these lattices, 
concerning lower and
upper limits for the critical temperatures, one may expect that these
bounds for the lower critical dimension, 
associated with a SG phase in the presence of
an external magnetic field, should apply to other lattices as well.

\begin{table}
\begin{center}
\begin{tabular}{||c|c|c||}			 \hline

Hierarchical & $k_{B}T_{c}/J$ 
& $k_{B}T_{c}/J$ \\
lattice & Uniform and Bimodal & Gaussian [$(\sigma/J)=1$] \\     \hline

Cell 1(a) & $0.880(1)$ & $0.661(1)$ \\

Cell 1(b) & $1.761(1)$ & $1.750(1)$ \\

Cell 1(c) & $0.980(1)$ & $0.891(1)$ \\ \hline
\end{tabular}
\end{center}
\caption{\small
The critical temperatures for $H_{0}=0$, with the fields following 
Eqs.~(\ref{eq:unifield})--(\ref{eq:gaussfield}), for the hierarchical 
lattices
defined by the cells of Fig.~\ref{fig:hlattices}. In the case of the
Gaussian distribution for the fields, we have chosen $(\sigma/J)=1$. The
error bars refer to the usual approach to criticality characteristic of the
RG technique.} 
\end{table}

\begin{table}
\begin{center}
\begin{tabular}{||c|c|c|c|c||}			 \hline

Hierarchical & Uniform & Bimodal 
& Gaussian [$(\sigma/J)=1$] & Gaussian ($H_{0}=0$)  \\     
lattice & $H_{0c}/J$ & $H_{0c}/J$ 
& $H_{0c}/J$ & $\sigma_{c}/J$  \\  \hline

Cell 1(a) & $0.530(1)$ & $3.919(1)$ & $0.433(1)$ & $3.036(1)$ \\

Cell 1(b) & $1.414(2)$ & $21.45(2)$ & $1.418(1)$ & $20.25(2)$   \\

Cell 1(c) & $0.590(2)$ & $5.907(2)$ & $0.566(2)$ & $5.884(2)$   \\
\hline
\end{tabular}
\end{center}
\caption{\small
Zero-temperature critical values of the fields following 
Eqs.~(\ref{eq:unifield})--(\ref{eq:gaussfield}), for the hierarchical 
lattices
defined by the cells of Fig.~\ref{fig:hlattices}. In the case of the
Gaussian distribution for the fields, we have chosen either
$(\sigma/J)=1$, or $H_{0}=0$. The
error bars refer to the usual approach to criticality characteristic of the
RG technique.} 
\end{table}

\begin{table}
\begin{center}
\begin{tabular}{||c|c|c||}			 \hline

Hierarchical &Exponent $y$ &Exponent $u$ \\
lattice & &  \\     \hline

Cell 1(a) & $0.245(8)$ & $0.140(8)$ \\

Cell 1(b) & $0.254(2)$ & $0.002(1)$ \\

Cell 1(c) & $0.223(2)$ & $0.007(3)$ \\ \hline
\end{tabular}
\end{center}
\caption{\small
The zero-temperature exponents $y$ and $u$ (defined in \eq{eq:stiffness}) 
for an initial uniform field $H_{0}$ in the range 
$0 \leq H_{0} \leq H_{0c}$, where $H_{0c}$ is given in Table~II for each
of the hierarchical lattices
defined by the cells of Fig.~\ref{fig:hlattices}.}
\end{table}

To conclude, we have investigated a nearest-neighbor-interaction 
Ising spin glass model, in the presence of an external magnetic field, 
on three different hierarchical lattices that approach the cubic lattice. 
In the beginning of the renormalization-group procedure, the magnetic 
field was considered as uniform, or randomly distributed, following 
either a bimodal or a Gaussian probability
distribution. In all cases considered, a spin-glass attractor was found, 
in the plane magnetic field versus temperature, which was associated with 
a low-temperature spin-glass phase. In the particular cases of 
hierarchical lattices for which the fractal dimension may be
changed easily by varying the number of parallel paths, we have verified
that the lower critical dimension, associated with a finite-temperature
spin-glass phase, lies in the interval $2.585<D_{l}<3$. 
The present results show that, in what
concerns the hierarchical lattices studied, 
{\em there is a spin-glass phase in
the presence of an external magnetic field}, contrary to the claim 
of recent numerical simulations on the cubic lattice 
\cite{young04,jorg}. 
Since the above results concern the existence of a
spin-glass attractor, under renormalization-group transformations, 
the question of replica-symmetry breaking
throughout this phase, which would characterize the critical lines presented
herein as Almeida-Thouless lines, represents a point that deserves further
investigations. 
Taking into account that hierarchical lattices have been a useful tool for
studying spin-glass systems, the present results motivate the
investigation of replica-symmetry breaking properties for 
spin glasses on hierarchical lattices, 
in the presence of external magnetic fields. It is possible that the
picture found previoulsy on the $D=3$ Migdal-Kadanoff hierarchical lattice,
without a magnetic field
\cite{drossel98,marinaricomment,drosselreply,drossel00}, exibiting
replica-symmetry breaking characteristics only close to the critical
temperature, may change under the presence of an external magnetic field.

\vskip 2\baselineskip

{\large\bf Acknowledgments}

\vskip \baselineskip
\noindent
We thank Prof. E.~M.~F. Curado for fruitful
conversations. The partial financial supports from
CNPq and Pronex/MCT/FAPERJ (Brazilian agencies) are acknowledged. 

\vskip 2\baselineskip

\end{document}